\documentclass[%
 aip,
 amsmath,amssymb,
reprint,%
]{revtex4-2}

\usepackage{graphicx}
\usepackage{dcolumn}
\usepackage{bm}

\usepackage[utf8]{inputenc}
\usepackage[T1]{fontenc}
\usepackage{mathptmx}
\usepackage{etoolbox}
\usepackage{xcolor} 
\usepackage{amsmath, gensymb} 
\usepackage[caption=false]{subfig}
\usepackage[colorinlistoftodos,prependcaption,textsize=small]{todonotes}
\usepackage{siunitx}
\usepackage{miller}
\usepackage{xr}
\usepackage[bookmarks=False]{hyperref}

\hypersetup{
    colorlinks=true,
    linkcolor=magenta,
    filecolor=magenta,      
    urlcolor=magenta
}

\makeatletter
\def\@email#1#2{%
 \endgroup
 \patchcmd{\titleblock@produce}
  {\frontmatter@RRAPformat}
  {\frontmatter@RRAPformat{\produce@RRAP{*#1\href{mailto:#2}{#2}}}\frontmatter@RRAPformat}
  {}{}
}%
\makeatother
\begin{document}

\newcommand{\AlOx}{Al$_2$O$_3$}
\newcommand{\PrUnits}{$\mu$C/cm$^2$}
\newcommand{\HZO}{Hf$_{0.5}$Zr$_{0.5}$O$_2$}
\newcommand{\HfO}{HfO$_2$}

\preprint{APL21-AR-09642}

\title[]{A multi-pulse wakeup scheme for on-chip operation of devices based on ferroelectric doped \HfO{} thin films} 

\author{S. Lancaster}
\email{suzanne.lancaster@namlab.com}
\affiliation{ 
NaMLab gGmbH, N\"{o}thnitzer Str. 64a, Dresden 01187, Germany
}
\author{T. Mikolajick}
\affiliation{ 
NaMLab gGmbH, N\"{o}thnitzer Str. 64a, Dresden 01187, Germany
}
\affiliation{Chair of Nanoelectronics, TU Dresden, N\"{o}thnitzer Str. 64, Dresden 01187, Germany}

\author{S. Slesazeck}
\affiliation{ 
NaMLab gGmbH, N\"{o}thnitzer Str. 64a, Dresden 01187, Germany
}

\date{\today}

\begin{abstract}
A wakeup scheme for ferroelectric thin \HZO{} films is presented, based on a gradual switching approach using multiple short pulses with a voltage amplitude roughly equal to the coercive voltage. This enables the on-chip wakeup and switching operation of ferroelectric devices such as tunnel junctions (FTJs) with identical pulses. After wakeup using alternating pulse trains which gradually switch the film polarization, FTJ operation is demonstrated to be as effective as after `normal' wakeup, with bipolar pulses of an amplitude larger than the coercive voltage.  In this case the voltage applied during wakeup was reduced by 26\%, thereby lowering the required operating power.
\end{abstract}

\maketitle


 \textbf{This article may be downloaded for personal use only. Any other use requires prior permission of the author and AIP Publishing. This article appeared in Appl. Phys. Lett. 120, 022901 (2022) and may be found at \hyperlink{https://doi.org/10.1063/5.0078106}{https://doi.org/10.1063/5.0078106}.} \\
 
 With the discovery of ferroelectricity in doped \HfO{} thin films \cite{boscke2011ferroelectricity}, the first fully CMOS-compatible ferroelectric (FE) was demonstrated. This has furthered the development of various FE devices, such as ferroelectric tunnel junctions (FTJs) \cite{ambriz2017complementary} or ferroelectric field effect transistors (FeFETs) \cite{trentzsch201628nm}. In particular, the integration of such devices into circuits such as artifical neural networks \cite{covi2021ferroelectric} has become more realistic, especially with the demonstration of back-end-of-line compatibility \cite{deshpande2021cmos, begon2021beol}. Nonetheless, some aspects of standard \HfO{} device characterization and measurement cannot be directly transferred to on-chip operation. \\

One such aspect is the wakeup effect which is common to \HfO-based FE thin films, whereby a film must undergo bipolar electric field cycling before optimal ferroelectric switching can be achieved. Typically, for a $\sim{}$ 10 nm \HZO{} (HZO) film, a maximum voltage V$_{max}$ of 4 V is needed for wakeup. However, with the addition of a dielectric (DE) interlayer such as that needed for bilayer FTJ operation, V$_{max}$ needed for wakeup and switching can be pushed up to \textgreater{} 6 V \cite{max2018ferroelectric}. \\

For on-chip operation of FE devices, lower operating voltages are preferable, since they reduce power consumption, as well as the need for integrating high voltage devices into the process technology. To that end, we here investigate a wakeup scheme where short, lower-voltage pulses gradually switch polarization from a pristine state. This is similar to an accumulative switching scheme \cite{mulaosmanovic2018accumulative}, whereby devices with single or few domains switch abruptly after the application of a defined number, \textit{n}, of non-switching pulses. In larger, multi-domain devices such as those featured here, that process is modified not only due to a statistically distributed coercive voltage of the domains \cite{saha2019phase}, but also potential inhomogeneities \cite{zhukov2010dynamics}, due to local electric fields or process variation. Thus switching is not abrupt, but happens stepwise as single domains switch on subsequent pulses, leading to a reproducible gradual switching behavior. This approach can be adapted in order to reduce the operation voltage of other FE devices down to within limits given by the available devices in a given process technology. This is a necessary step in the integration of FE devices on-chip which further allows identical pulses to be used for wakeup as well as device operation, which is desirable for on-chip applications \cite{yu2018neuro}. \\

Bottom electrodes were prepared with 12 nm of TiN deposited under UHV conditions. HZO and \AlOx{} layers of 12 and 2 nm, respectively, were deposited via atomic layer deposition in an Oxford OpAL system. 12nm top TiN electrodes were sputtered, followed by a 20 s crystallization anneal at 600 $\degree$C to achieve ferroelectricity. Finally, capacitor structures were deposited by evaporating 10/25 nm Ti/Pt through a shadow mask, and the top TiN was etched away with SC-1. \\

Electrical measurements were carried out using PMUs controlled by Keithley 4225 RPMs connected to a Keithley 4200 parameter analyzer. First, the switching kinetics of the FTJ stacks were measured. Next, gradual switching (also known as analog switching) was investigated, i.e. how effectively the polarization state could be changed, stepwise, via the application of multiple pulses. Finally, gradual switching was applied with alternating polarities to cycle devices from a pristine state, herein referred to as `multi-pulse wakeup'. \\

\begin{figure}[!ht]
    \centering
    \subfloat[]{
        \includegraphics[height=3.2cm]{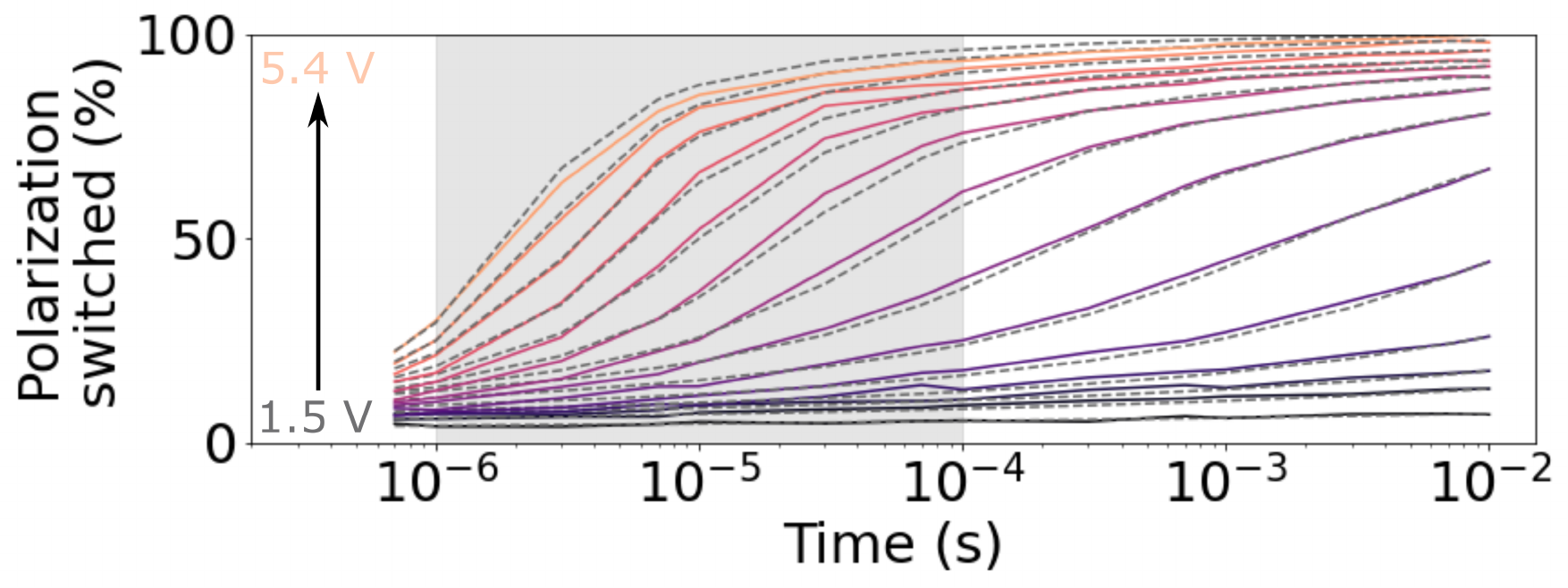}
        \label{Switching_On}
    } \\
    \subfloat[]{
        \includegraphics[height=3.2cm]{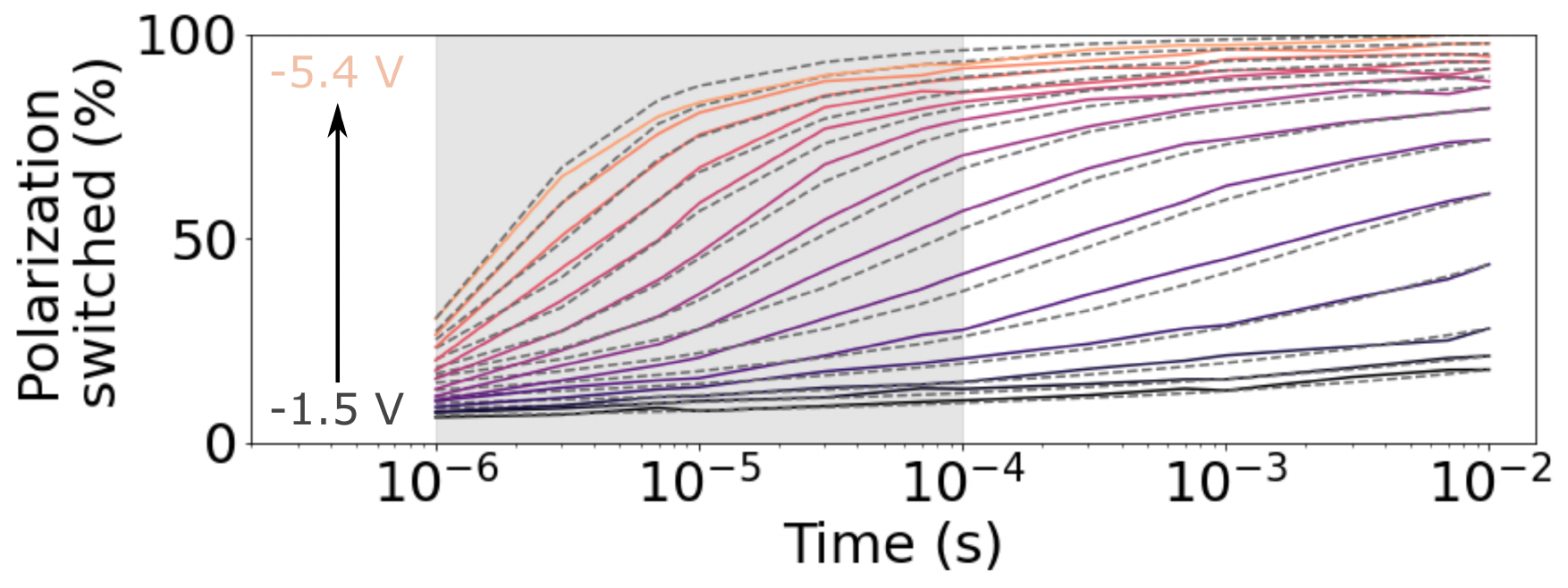}
        \label{Switching_Off}
    } \\
    \subfloat[]{
        \includegraphics[height=3cm]{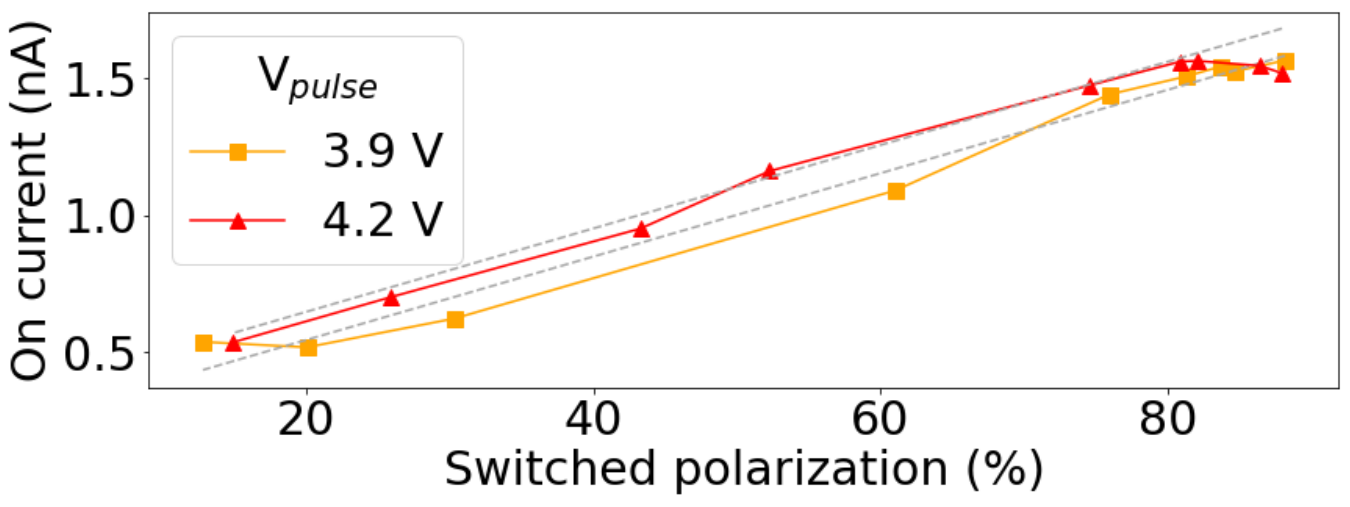}
        \label{Pol_Current}
    }
    \caption{Switching kinetics of FTJ devices: amount of switched polarization with pulses of different voltages (0.3 V increments) and pulse widths, for \protect\subref{Switching_On} switching into the $P_{up}$ (low-resistance) state, and \protect\subref{Switching_Off} into the $P_{down}$ (high-resistance) state. \protect\subref{Pol_Current} The FTJ on-current (after wakeup) shows a linear correlation with the amount of switched polarization, as demonstrated when switching into P$_{up}$ at two different switching voltages.}
    \label{Switching}
\end{figure}

Switching kinetics measurements were performed on the bilayer films, where pulses of increasing width t$_{pulse}$ (0.5 $\mu$s - 10 ms) and amplitude V$_{pulse}$ ($\pm$ 1.5 - 5.4 V) are applied and the amount of switched polarization is quantified \cite{mulaosmanovic2020interplay, materano2020polarization}. This is done by using a pulse of opposite polarity, followed by a second pulse to fully account for non-switching current contributions. The switching kinetics of these FTJ devices are shown in figure \ref{Switching_On} for switching into the P$_{up}$ state and figure \ref{Switching_Off} for switching into the P$_{down}$ state. It can be seen that the switching is asymmetric from off- to on-state and vice versa, with an easier switching into the off-state. This asymmetry is due to differences in the coercive fields and switching distributions in each polarity, which stem from the inherent asymmetry of the FTJ stack and generation of an internal bias field due to fixed charges \cite{park2015study}, charges trapped during operation \cite{fontanini2021polarization}, or work-function difference of the electrodes. \\

The data can be well fitted using a nucleation-limited switching (NLS) model \cite{tagantsev2002non, boyn2017learning} (grey dashes), where the amount of switched polarization S follows a lorentzian distribution centered on a mean switching time $t_{mean}$ for each voltage V:
\begin{equation}
S_{\pm} (t, V) = \frac{1}{2} \mp \frac{1}{\pi}arctan\frac{log(t_{mean}(V)) - log(t)}{\Gamma (V)}
\end{equation} \\

where $\Gamma$ is the Lorentzian linewidth and t is the switching time. Furthermore, by measuring the current through the FTJ after each applied pulse (when switching into the $P_{up}$, or low-resistive state), it can be shown that the amount of switched polarization correlates linearly with the on-current. Thus, in this paper, we perform device characterization using the metric of switched polarization, as this simplifies the measurement routine, with the knowledge that the conclusions drawn can also be applied to FTJ on-current. \\

Standard FE switching is usually performed at a voltage of $\thicksim 150\%$ or more of the coercive voltage V$_c$ for a given pulse width. In contrast, analog switching is the gradual switching of the polarization via the application of multiple fast pulses below this amplitude \cite{mulaosmanovic2018accumulative}. The primary basis of analog switching is the nucleation and growth of FE domains on subsequent pulses \cite{saha2019phase}, thus the NLS in our FE/DE bilayers proves that analog switching is viable, despite internal bias fields arising from the stack asymmetry. Recent work has highlighted the role of charge injection in polarization switching, especially in the case of thicker interfacial layers \cite{park2021polarizing}. For that reason, an additional contribution of switching due to charges accumulated on subsequent pulses, thereby altering the effective field in the FE layer, can not be ruled out. In this case one would expect a further impact of the additional dielectric layer on both the charge trapping and field distribution. \\

From the plots in figure \ref{Switching}, we can estimate the pulse parameters which are most suitable for gradual switching of the devices. To perform these experiments, the device was prepoled into the $P_{down}$ state, before switching into the $P_{up}$ state with \textit{n} pulses of the chosen pulse parameters, with a delay between pulses set to t$_{del}$ = 1 $\mu$s. Switching was quantified as for the previous experiment (see inset). Full switching was measured with triangular pulses of 6.5 V, 500 $\mu$s and reset was performed at 6.5 V, 100 kHz for 10 cycles. \\

\begin{figure}[!hb]
    \centering
    \subfloat[]{
        \includegraphics[height=3.2cm]{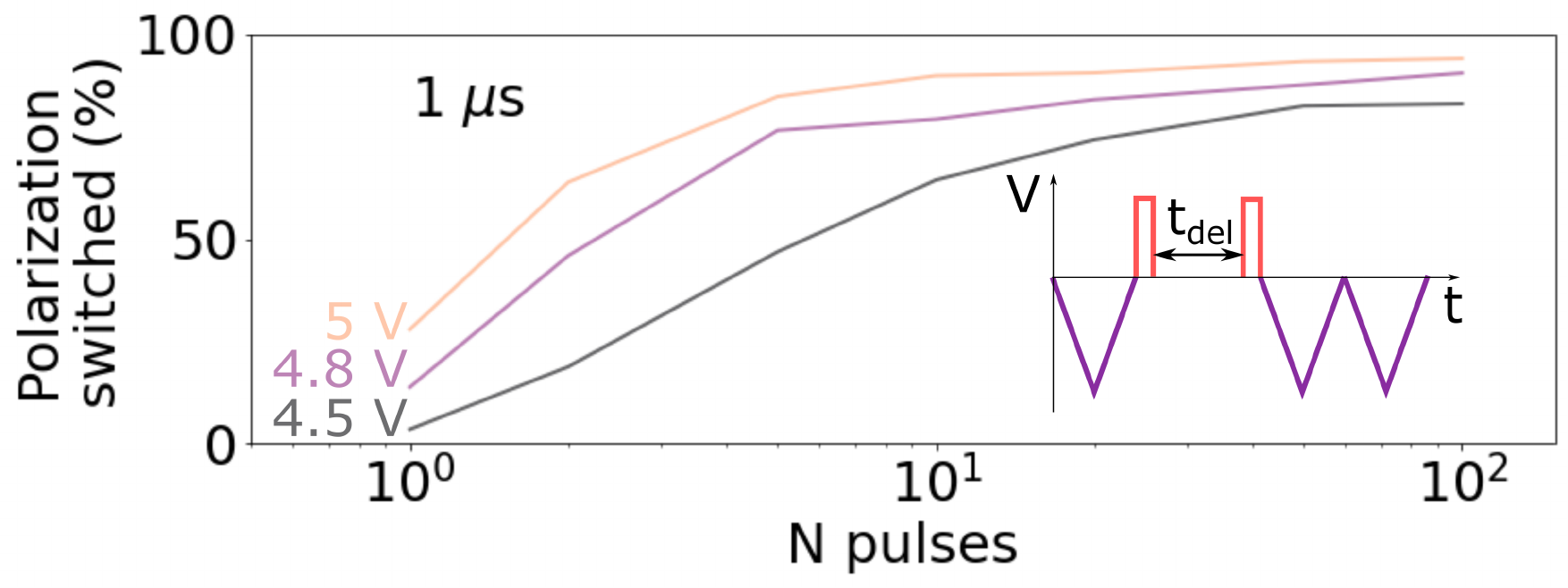}
        \label{Accumulative_1us}
    } \\
    \subfloat[]{
     \includegraphics[height=3.2cm]{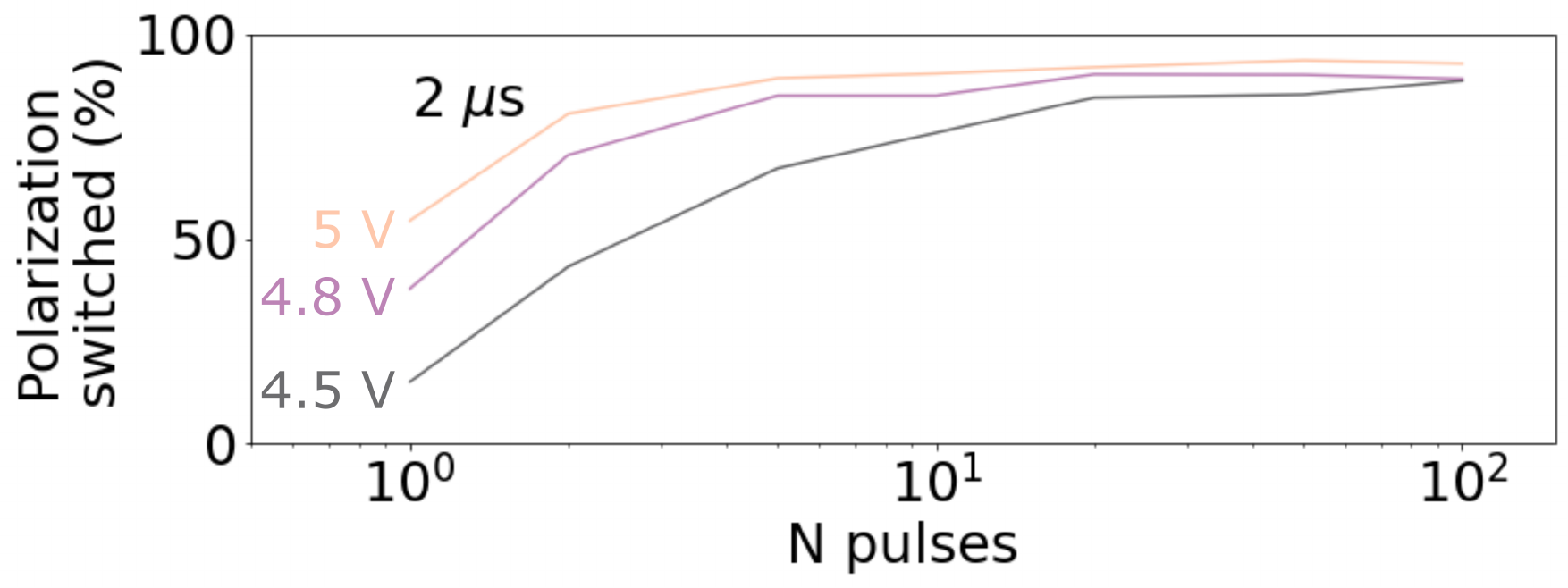}
        \label{Accumulative_2us}
    } 
    \caption{Percentage of polarization switched when applying $n$ square pulses at different voltage amplitudes for \protect\subref{Accumulative_1us} 1 $\mu$s width or \protect\subref{Accumulative_2us} 2 $\mu$s width, for switching into the $P_{up}$ state. Inset in \protect\subref{Accumulative_1us} shows the applied pulse train for performing the measurement.}
    \label{Accumulative}
\end{figure}

In this experiment we pick 1 or 2 $\mu$s pulses, as these offer a chance for switching over a large range of polarization in 100 pulses with amplitudes below 5 V, in both directions (grey shaded regions in figure \ref{Switching}). For example, a single 1 $\mu$s pulse at 4.5 V switches $\sim{}$17\% of the polarization, while a 100 $\mu$s pulse at 4.5 V switches $\sim{}$87\%. In figures \ref{Accumulative_1us} and (b), we can see that with different pulse parameters a polarization range up to $\sim{}$70\% can be covered (i.e. from 1 pulse to 100). Moreover, the saturation polarization also depends on the pulse parameters, mainly related to the rate of domain nucleation \cite{saha2019phase}. Thus both the range of polarization covered by the gradual switching scheme, and the saturation value of the polarization, can be controlled via the pulse parameters. \\

Further gradual switching experiments are shown in the supplementary (figure S1), where it can be seen that the voltage reduction has a limit, beyond which even long pulses of 100 $\mu$s do not lead to a high switching efficiency. \\

Based on these results, a wakeup scheme is proposed where instead of bipolar individual pulses, multiple pulses close to or below the coercive voltage are applied subsequently first in one polarity, and then the other. Through this approach, the polarization is gradually switched in each opposing direction. This is repeated over many cycles, resulting in a full wakeup of the device. With the aim of reaching a wide range of polarization switching, pulses of t$_{pls}$ = 1 $\mu$s and a voltage of 4.8 V were chosen, with \textit{n} = 20, beyond which the polarization is seen to saturate. This means that the voltage used for wakeup could be reduced by 26\%. \\

Since the delay time between pulses should also influence the eventual saturation polarization  \cite{saha2019phase}, and there should be no frequency limitation of on-chip wakeup, we chose a time delay of 1 $\mu$s, leading to an effective cycling frequency of 12.5 kHz (including all pulses in both polarities).  \\

\begin{figure}[!th]
    \centering
    \subfloat[]{
        \includegraphics[height=2.3cm]{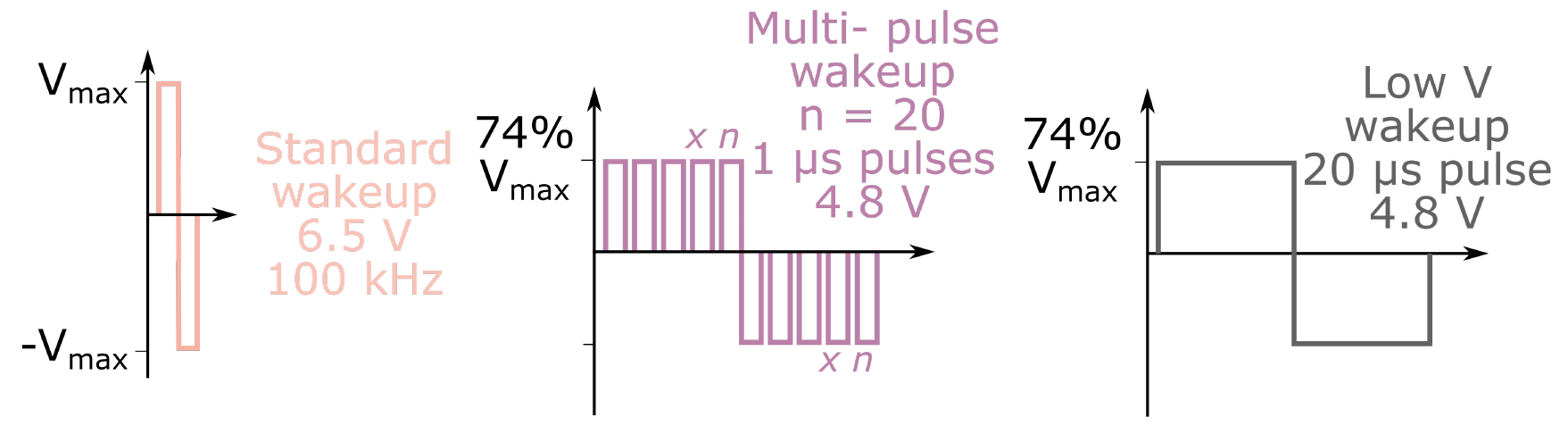}
        \label{FTJ_pls}
    } \\
     \subfloat[]{
        \includegraphics[height=4.4cm]{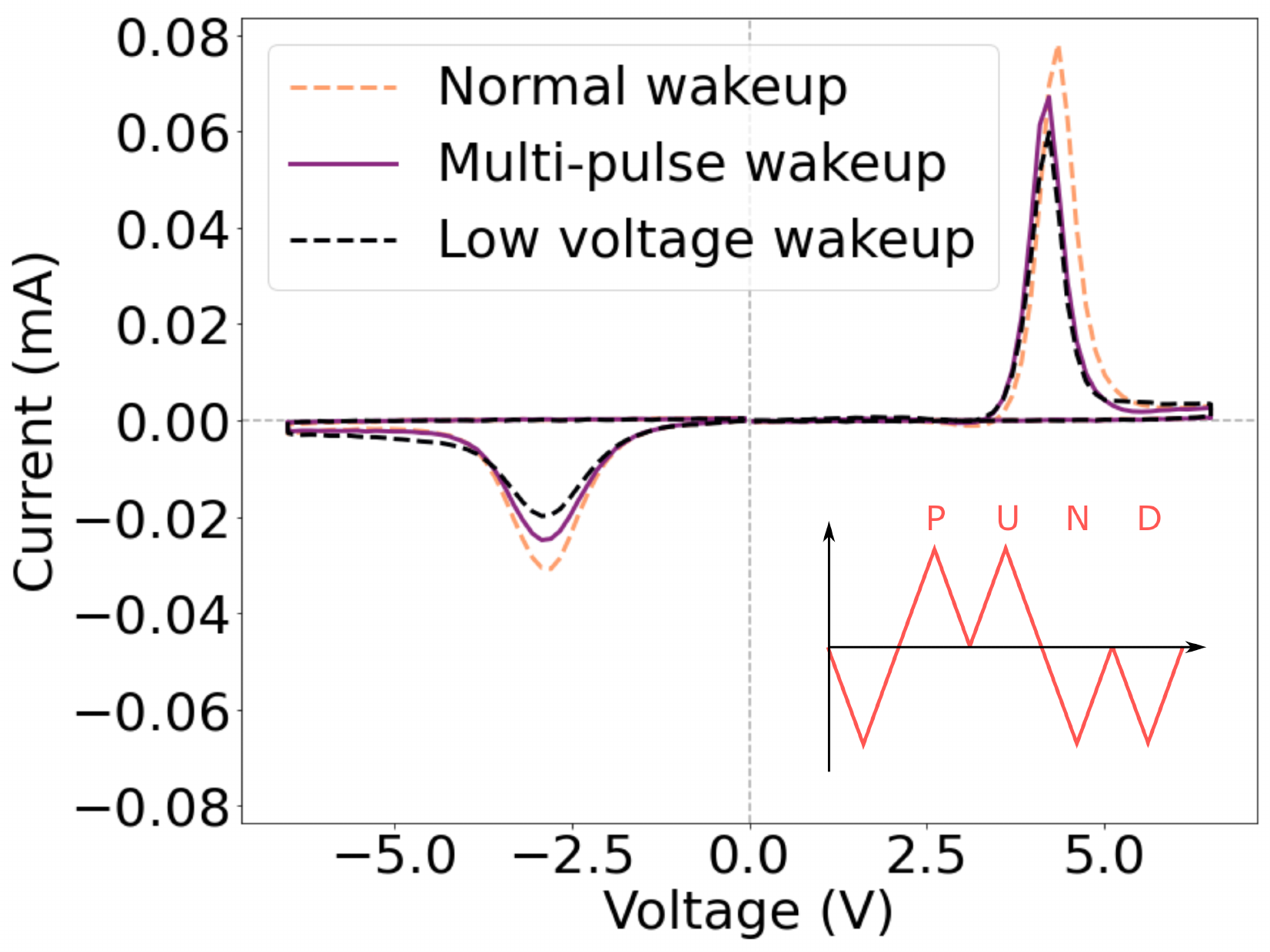}
        \label{PUND_IV}
    } \\
    \subfloat[]{
        \includegraphics[height=4.4cm]{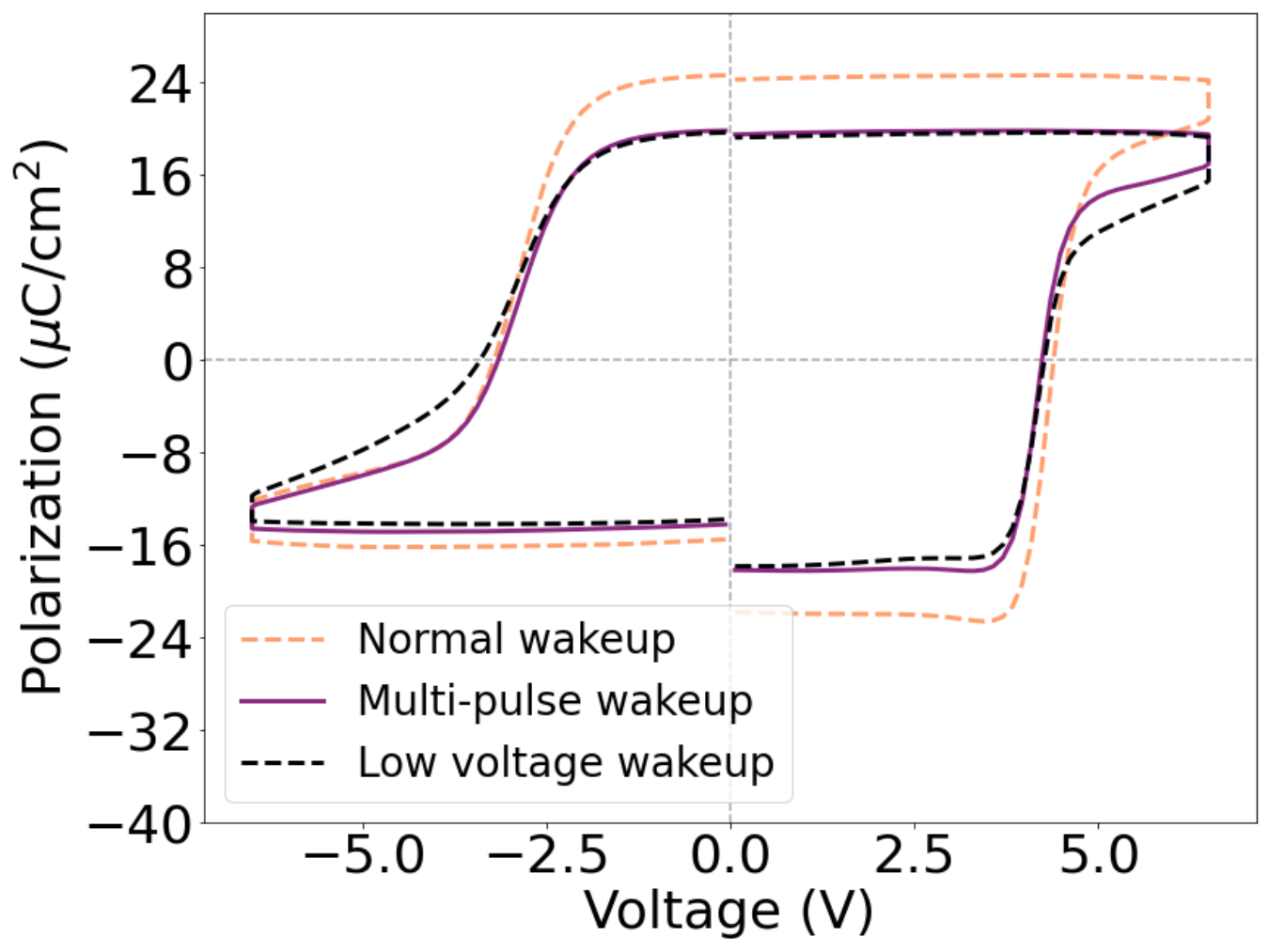}
        \label{PUND_PV}
    } \\
    \subfloat[]{
        \includegraphics[height=4.7cm]{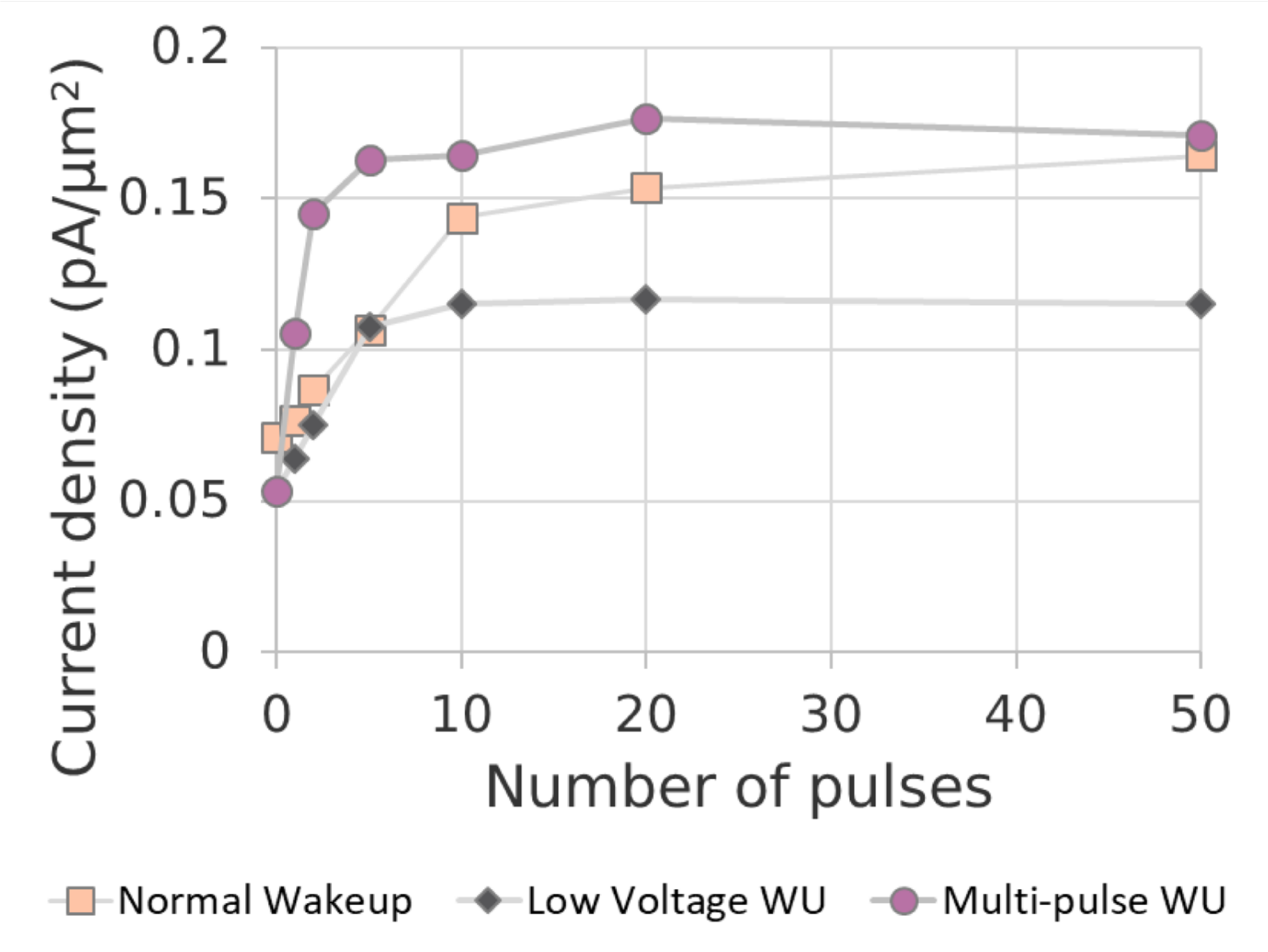}
        \label{FTJ_msr}
    } 
    \caption{\protect\subref{FTJ_pls} Schematics of different wakeup pulse schemes. \protect\subref{PUND_IV} I-V switching characteristics measured at 1 kHz after 1e$^3$ cycles wakeup under each scheme, \protect\subref{PUND_PV} corresponding P-V curves, and \protect\subref{FTJ_msr} FTJ On-current measured after different wakeup schemes and for different numbers of switching pulses}
    \label{FTJ}
\end{figure}

Analog FTJ operation after multi-pulse wakeup was compared to devices from the same sample woken up under standard conditions, i.e. bipolar 6.5 V pulses at 100 kHz. Another wakeup scheme was made for comparison, which consisted of bipolar 4.8 V pulses with $t_{tot} = n\cdot t_{pls}$. Each of these three wakeup schemes (depicted in figure \ref{FTJ_pls}) was applied to a fresh device for 1e$^3$ cycles.  \\

On the woken-up devices, switching characteristics were measured (figures \ref{PUND_IV} and \ref{PUND_PV}) using a PUND (Positive Up Negative Down) waveform, shown inset. In these measurements, plotting the difference in current of pulses P and U in positive polarity, and of N and D in negative polarity, give the pure switching currents. In the PUND measurements, a slight reduction in the coercive field is observed after both multi-pulse and low-voltage wakeup, since domains with the highest coercive fields are not switched by this method. \\

When integrating the switching current measured after multi-pulse wakeup, 82\% of domains are found to contribute to switching (as compared to 'full switching' measured after standard wakeup). From the gradual switching experiments shown in figure \ref{Accumulative_1us}, we find that a gradual switching scheme using the same parameters (20x 1 $\mu s$ pulses at 4.8 V) switches 84\% of domains, compared to full switching. This implies that the same domains contribute in the gradual switching regime, whether a normal or multi-pulse wakeup is first applied to the device. In other words, a multi-pulse wakeup is expected to be as effective as wakeup at higher voltage, when a device is being operated using identical pulses, as domains with a higher coercive field are excluded from the switching behaviour in both cases. Finally, it should be noted that applying the lower voltage (4.8 V) for wakeup at a standard frequency (100 kHz) results in a drastically reduced switching behaviour, as shown in the supplementary (figure S2). Thus the gradual switching behaviour of the multi-pulse wakeup is clearly effective at increasing the amount of woken-up domains, even at lower voltage. \\

In order to assess the suitability of the multi-pulse wakeup scheme for device operation, FTJ devices were analyzed using an analog switching experiment. As in figure \ref{Accumulative}, $n$ number of switching pulses were applied to the woken-up FTJs, and this time the read-out was the FTJ on-current at 2 V. The on-current vs. number of switching pulses is shown in figure \ref{FTJ_msr} after different wakeup regimes. Notably, when the same identical pulses are used for device operation as for multi-pulse wakeup, the devices perform as well as if they were woken up with the standard scheme. We see strikingly comparable behavior in the maximum tunneling electroresistance (TER, $\sim{}$ 3.5) and on-current (I$_{on}$, 0.18 pA/$\mu m^2$). A steeper dependence on pulse number is measured after multi-pulse wakeup, which fits with a lower coercive voltage observed in switching experiments performed after wakeup, for both the low-voltage and multi-pulse wakeup schemes. Additional multi-pulse wakeup schemes were explored in supplementary figure S3. \\

These results indicate that the multi-pulse wakeup scheme is sufficient in activating all the domains which contribute to analog FTJ operation. All domains which are switchable via an analog switching scheme with identical pulses are accessed via multi-pulse wakeup, and any other domains (i.e. those with the highest coercive fields) cannot contribute to the analog switching behaviour, unless the device is set with increasing pulse amplitudes or widths \cite{begon2021beol}, possibly necessitating additional high-voltage devices on-chip. Additionally, it shows that the polarization can be gradually switched with identical pulses even from the pristine state. This is notable since gradual switching occurs due to domain dynamics on sequential pulses \cite{boyn2017learning, saha2019phase}, which are likely to vary between pristine and woken-up HZO. \\

In the case of the longer, low-voltage wakeup pulses, the device behavior is less efficient, with a lower saturation I$_{on}$ and thus a lower overall TER. This was confirmed for different pulse schemes, as shown in the supplementary (figure S4). While the IV curves of the samples woken up with either a pulsed wakeup or a low-voltage wakeup are nominally similar (figures \ref{PUND_IV} and \ref{PUND_PV}), small differences in the switching behaviour can be seen. Notably, the switching IV after low-voltage wakeup (black dashed line) shows the delayed switching of some domains at high voltages, pointing to a pinning of domains \cite{pesic2016impact}, leading to a non-negligible impact on the FTJ behavior. It was also shown previously \cite{max2019direct} that both on-current and TER increase with wakeup cycling; thus, one might expect that the effective cycling applied under multi-pulse wakeup is higher due to the additional effect of voltage ramping on each short pulse. \\

To conclude, we have demonstrated that a multi-pulse wakeup scheme, based on the analog switching behaviour of FE domains, can achieve the same device operation in bilayer FTJs as a standard wakeup scheme at a higher voltage. This solves two problems for the on-chip integration of FE devices: firstly, it reduces the necessary operating voltage, in this case by 26\%, and pulse width. This reduces power consumption for supporting circuitry and enables operation at a lower voltage, and can be applied to any FE device to reduce the operation voltage. Secondly, it allows for identical pulses to be applied both during the wakeup and during the on-chip device set/reset operations. This further simplifies the device requirements for the embedding technology. \\

 Given that the 2 nm DE layer in these samples could lead to considerable domain relaxation between subsequent pulses, one could expect that the multi-pulse wakeup could be even more efficient in layers of HZO only, or with thin oxide layers, where all or most domains can be reached with a gradual switching scheme  \cite{mulaosmanovic2020investigation}. Finally, wakeup and device operation using fast pulses could have other benefits not investigated here, such as a better control over charge trapping at the FE/DE interface. 

\section*{Supplementary Material}
See Supplementary Material for comparisons of gradual switching and FTJ operation using additional pulse parameters, efficiency of a 100 kHz wakeup scheme at 4.8 V, and additional data on multi-pulse and low-voltage wakeup.

\begin{acknowledgments}
S.L. acknowledges funding from the FLAG-ERA JTC 2019 grant SOgraphMEM through the DFG (MI 1247/18-1). The authors further acknowledge funding by the European Commission through the BeFerroSynaptic project (GA:871737). FTJ samples were provided by Dr. Benjamin Max from TU Dresden. The authors acknowledge Dr. Erika Covi for her valuable input on circuit requirements. 
\end{acknowledgments}

\section*{Data Availability Statement}
The data that support the findings of this study are available from the corresponding author upon reasonable request.

\nocite{*}
\bibliography{pulsed_wakeup}
\end{document}


\widetext
\begin{center}
\textbf{\large Supplementary Material: A multi-pulse wakeup scheme for on-chip operation of devices based on ferroelectric doped \HfO{} thin films}
\end{center}
\setcounter{equation}{0}
\setcounter{figure}{0}
\setcounter{table}{0}
\setcounter{page}{1}
\makeatletter
\renewcommand{\thefigure}{S\arabic{figure}}

\begin{section}{Gradual switching experiments on FTJ samples}

\begin{figure}[h]
    \centering
    \subfloat[]{
        \includegraphics[height=3.2cm]{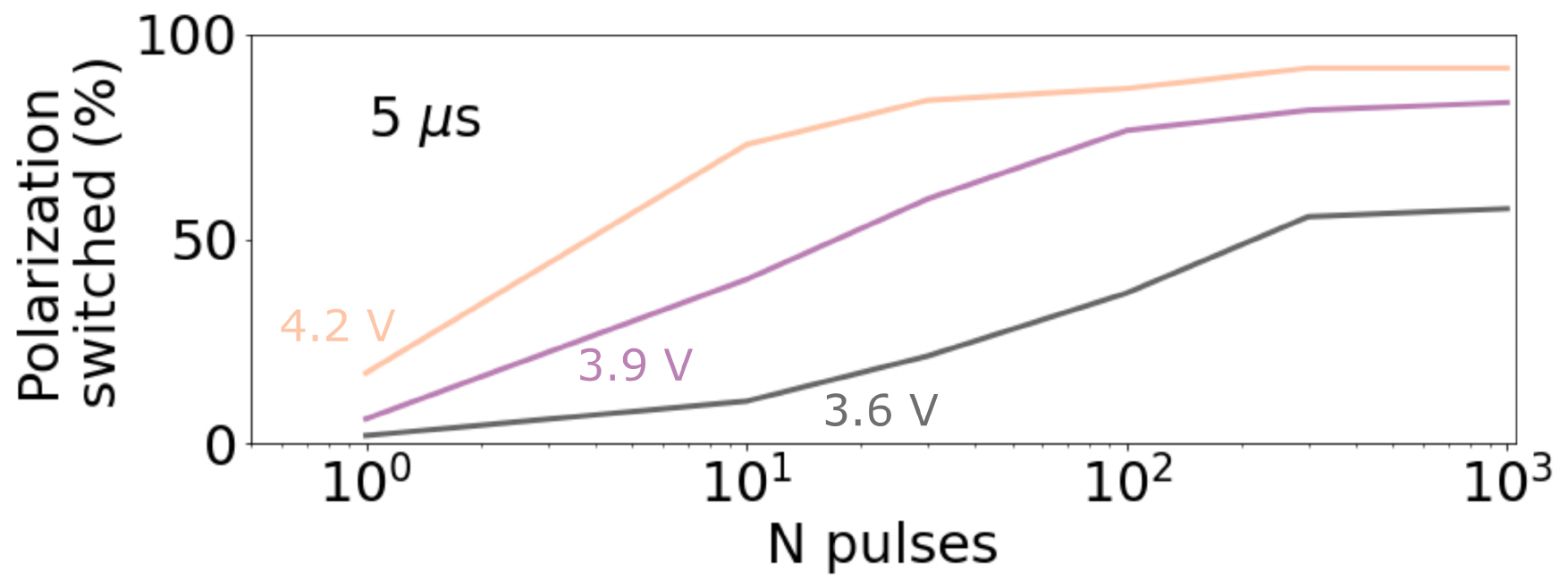}
        \label{acc_5us}
    } \\
     \subfloat[]{
        \includegraphics[height=3.2cm]{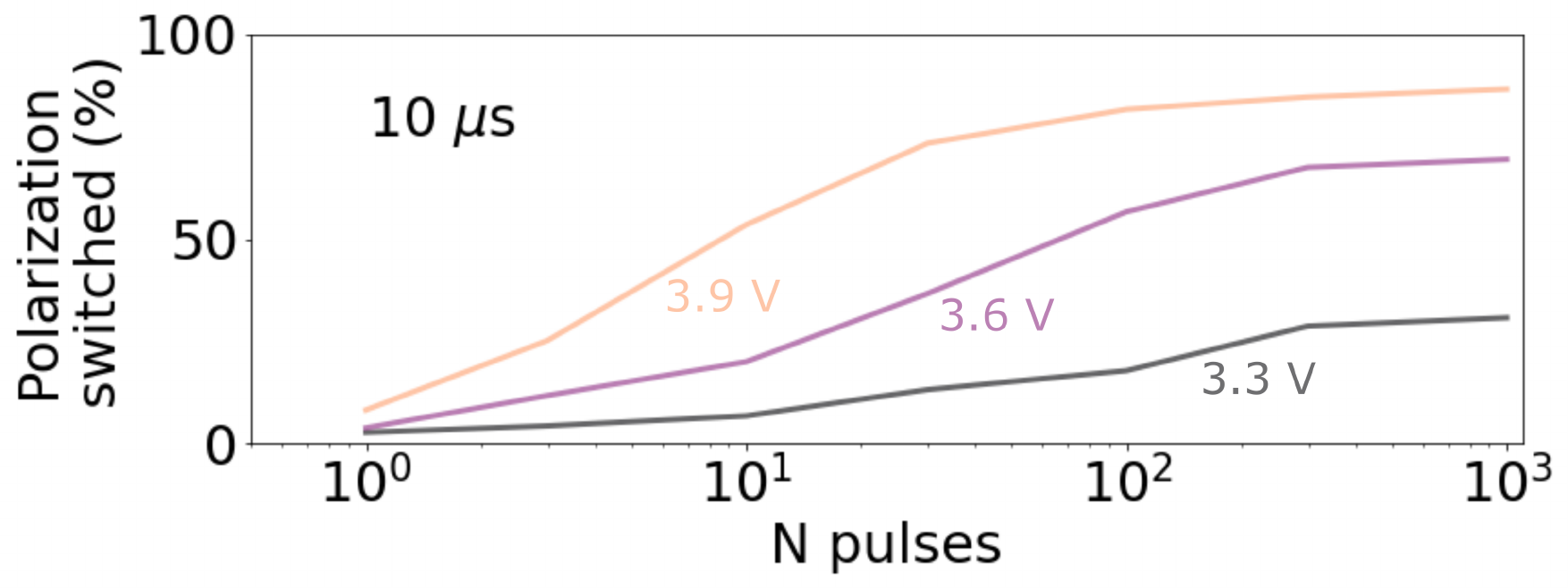}
        \label{acc_10us}
    } \\
    \subfloat[]{
        \includegraphics[height=3.2cm]{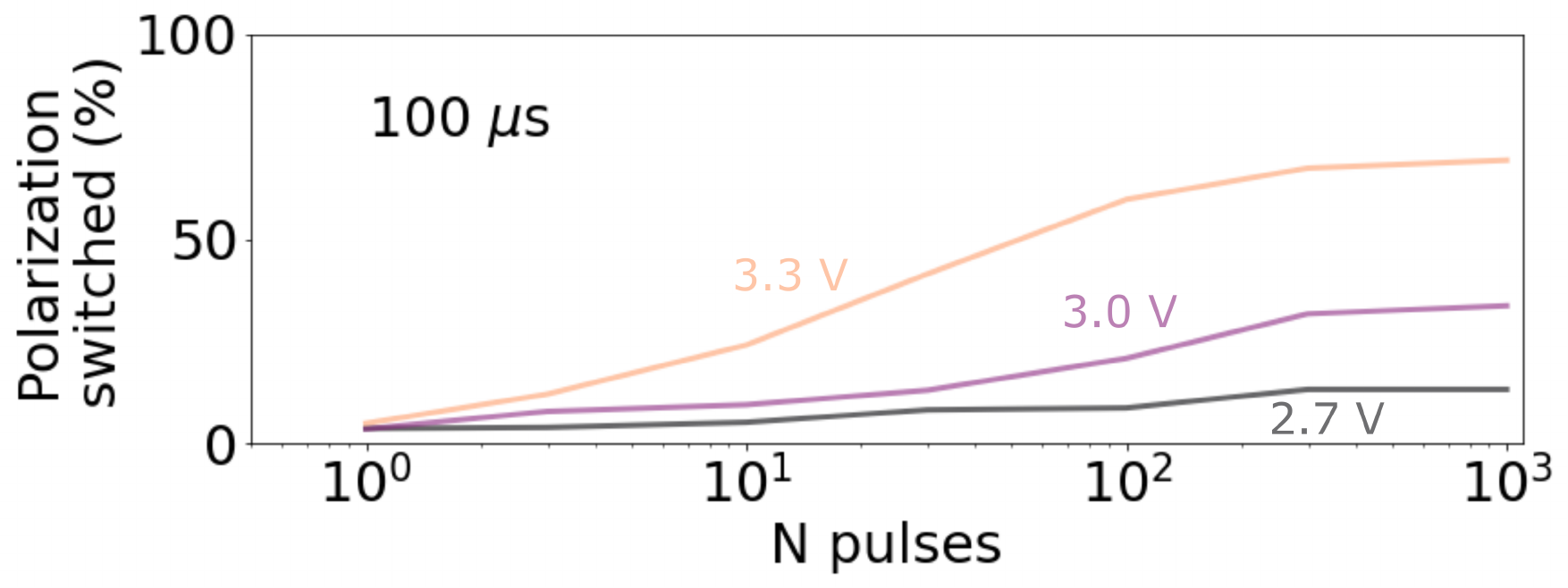}
        \label{acc_100us}
    } 
    \caption{Gradual switching experiments at various voltages, up to 1e$^3$ pulses, for pulse widths of: \protect\subref{acc_5us} 5 $\mu$s, \protect\subref{acc_10us} 10 $\mu$s and \protect\subref{acc_100us} 100 $\mu$s. The delay time between each subsequent pulse was kept at 1$\mu$ s for each experiment.}
    \label{acc_supplementary}
\end{figure}

Figure \ref{acc_supplementary} shows the amount of polarization switched after \textit{n} pulses, for different pulse widths and voltages. It is clear that as the voltage is reduced too far below the coercive voltage (as measured at 1 kHz), the total amount of switched polarization reduces, even after 1000 cumulative switching pulses. As the voltage is further reduced, even moving to relatively long pulse times (100 $\mu$s pulses, figure \ref{acc_100us}) doesn't allow for a large percentage of the polarization to switch. From this one can conclude that the optimal parameters for using multi-pulse wakeup to switch a large amount of domains is to use a voltage less than the coercive voltage, but where a single pulse still leads to some switching. From there, one can further vary the pulse width or number of pulses, to best fit the application. It should be noted that while many non-switching pulses should eventually switch an HZO film, as in accumulative switching [1], in practice and in devices with an additional DE layer, domain relaxation probably leads to lower saturation polarization [2]. 
\end{section}

\begin{section}{Low voltage, high frequency wakeup}
\begin{figure}[!h]
    \centering
    \subfloat[]{
        \includegraphics[height=5cm]{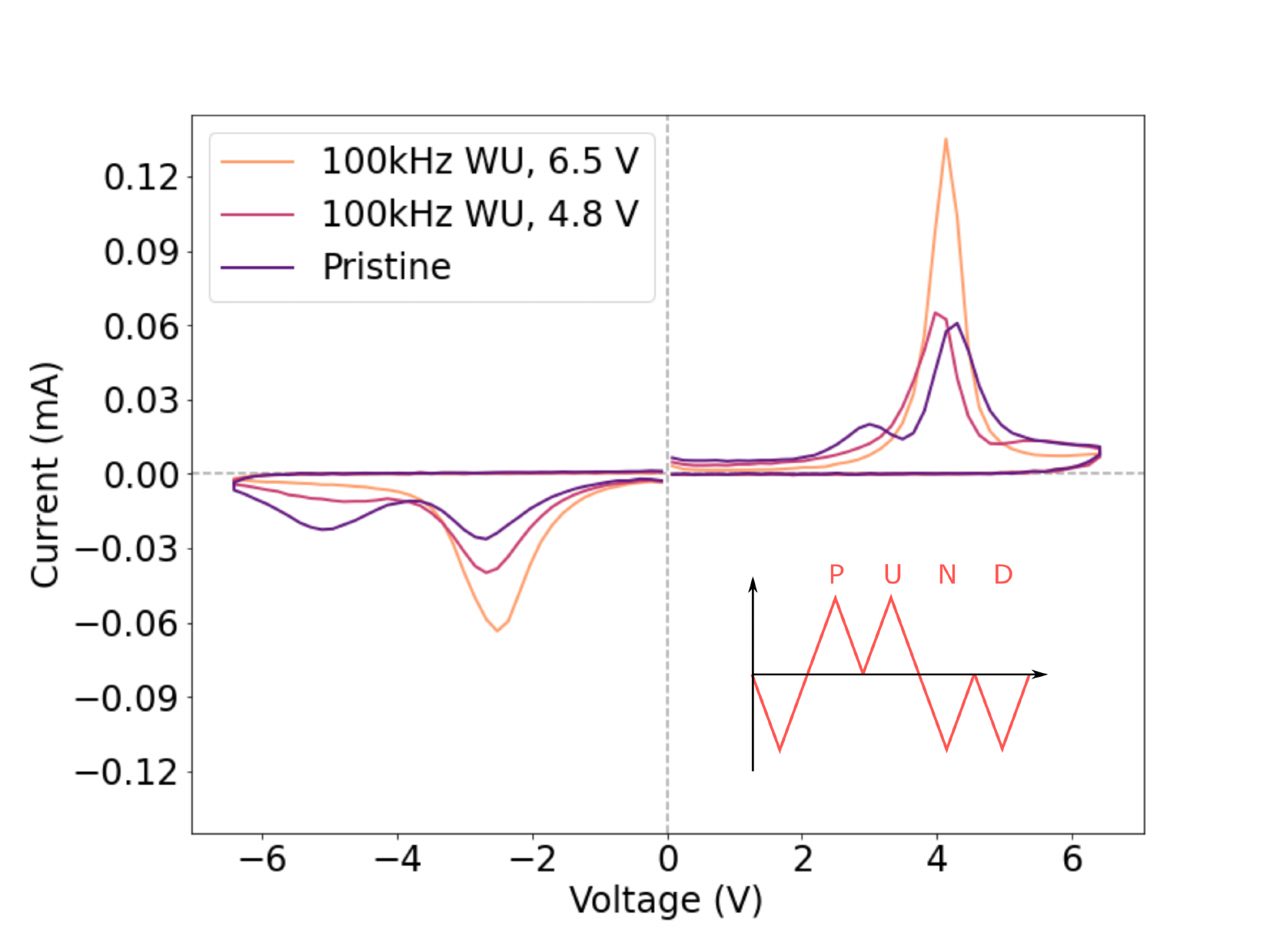}
        \label{PUND_IV_Standard}
    } 
    \subfloat[]{
        \includegraphics[height=5cm]{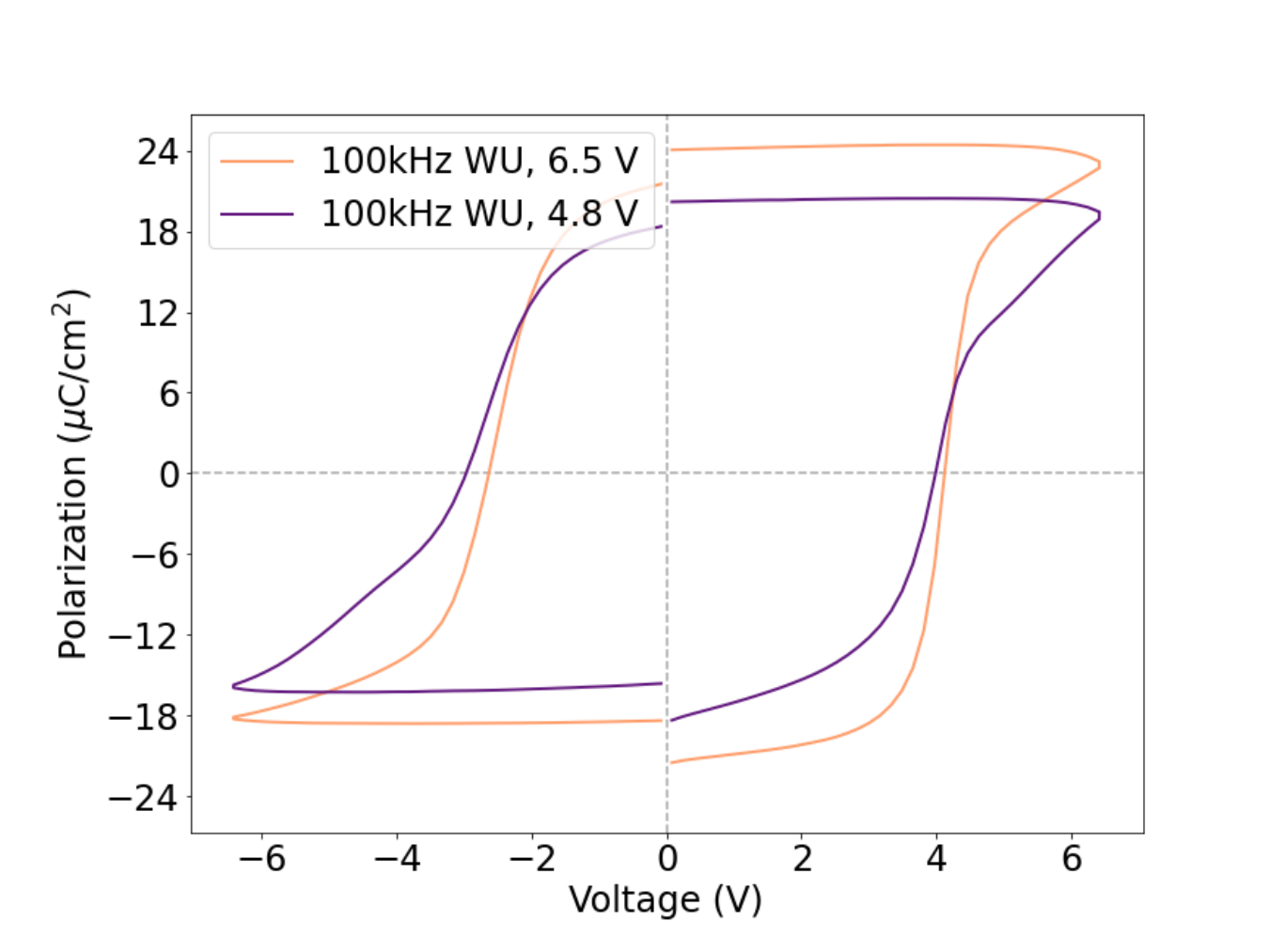}
        \label{PUND_PV_Standard}
    } \\
    \subfloat[]{
        \includegraphics[height=5.2cm]{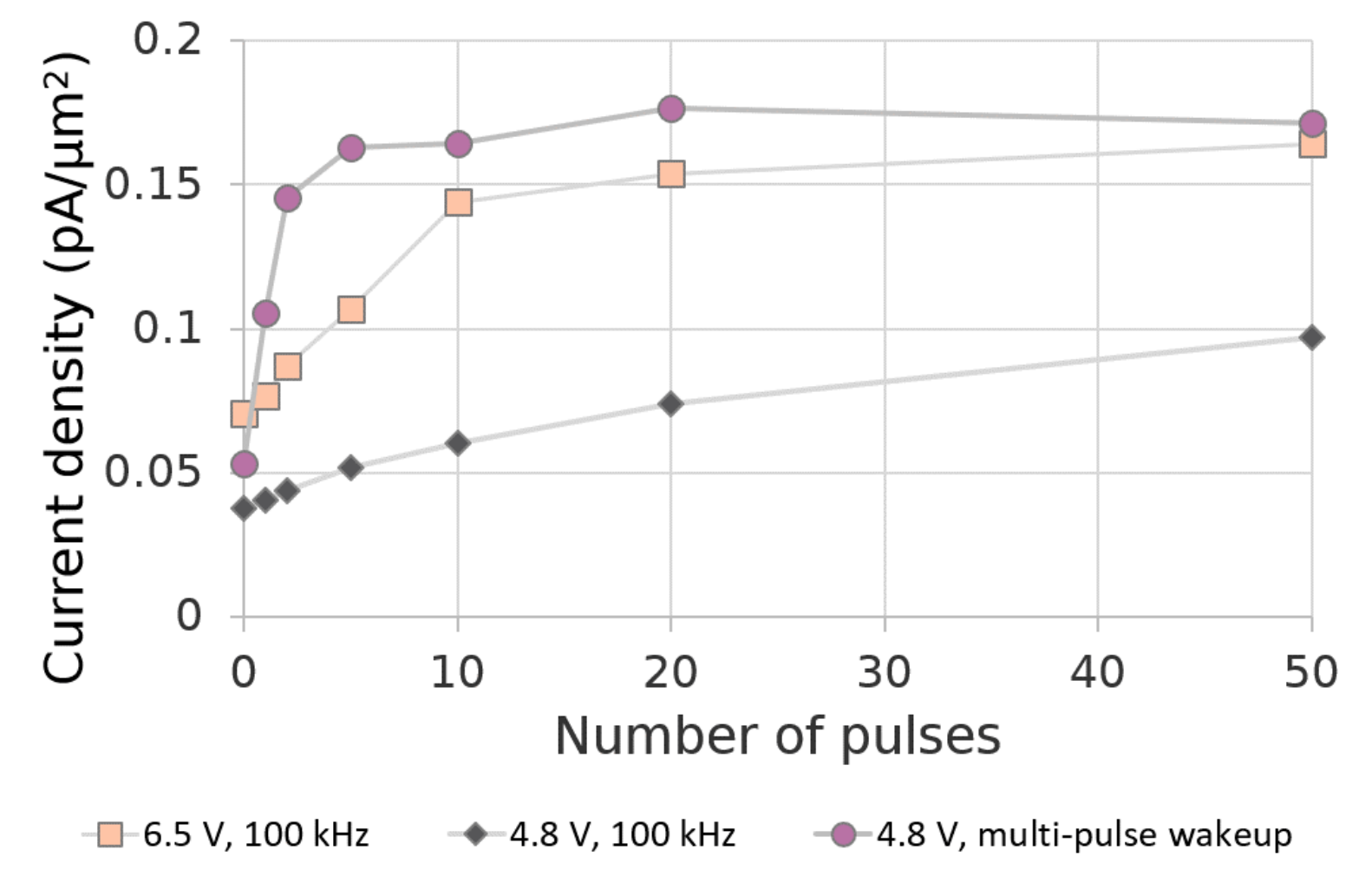}
        \label{FTJ_msr_Standard}
    } 
    \caption{\protect\subref{PUND_IV_Standard} I-V switching characteristics measured at 1 kHz after 1e$^3$ cycles wakeup under pulsed wakeup, and 100 kHz wakeup at 4.8 or 6.5V; \protect\subref{PUND_PV_Standard} corresponding P-V curves, and \protect\subref{FTJ_msr_Standard} FTJ On-current measured after each wakeup scheme, and for different numbers of switching pulses}
    \label{Standard}
\end{figure}

To prove the efficacy of the pulsed wakeup, the 'standard' wakeup scheme (i.e. 100 kHz pulses, or bipolar pulses with 5 $\mu$s in each polarity) was reduced to 4.8 V. In figures \ref{PUND_IV_Standard} and \ref{PUND_PV_Standard}, the switching characteristics of an FTJ stack after such a wakeup are compared to those after wakeup at a higher voltage of 6.5 V. It is clear that many more domains are woken up by using the higher voltage. As shown in the main text, when instead using the lower voltage of 4.8 V but applying a gradual switching or multi-pulse wakeup, where many identical pulses are applied first in one polarity and then the other, an intermediate number of domains can be switched. The device operation is then comparable to a device woken up with higher voltage, 100 kHz pulses. As is shown in figure \ref{FTJ_msr_Standard}, in the case when the 100 kHz wakeup is applied at a reduced voltage of 4.8 V, both the on-current and TER are reduced. This is to be expected, as not all of the domains which should be switched during the device operation have been properly woken up by this pulse scheme. 
\end{section}

\begin{section}{Additional multi-pulse wakeup schemes}
\begin{figure}[h]
\subfloat[]{
    \includegraphics[height=6cm]{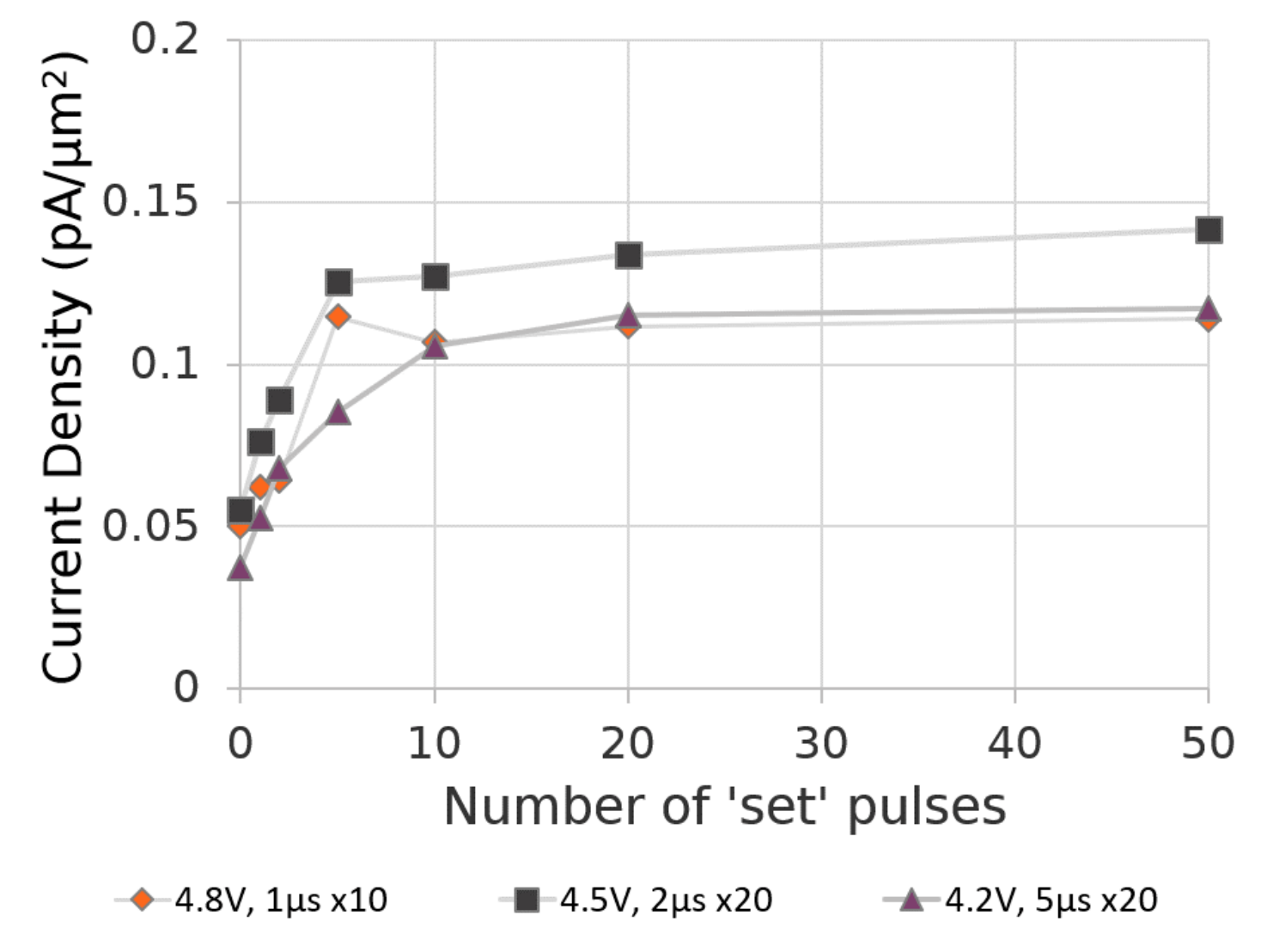}
    \label{FTJ_PWU}
    } \\
    \subfloat[]{
    \includegraphics[height=3cm]{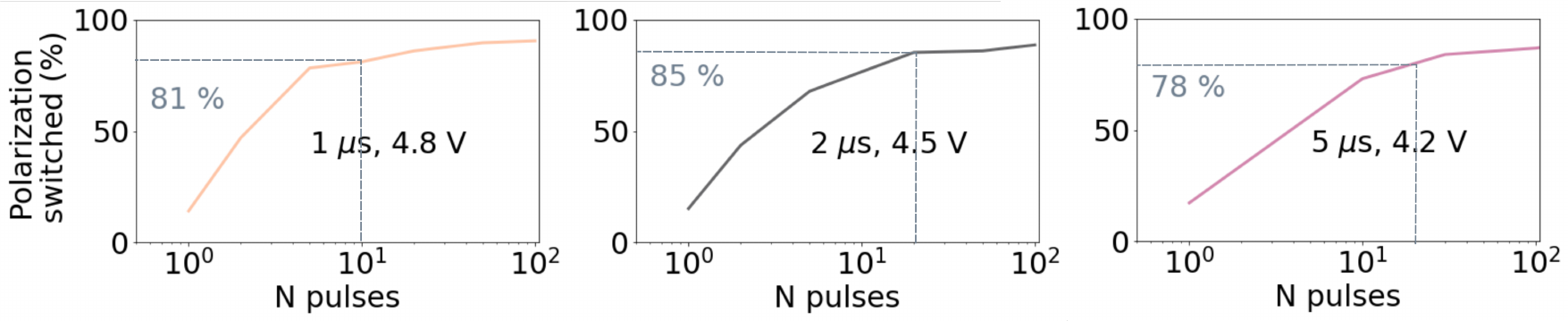}
    \label{PWU_Gradual}
    }
    \caption{\protect\subref{FTJ_PWU} FTJ On-current measured after different pulsed wakeup schemes: (normal wakeup, multi-pulse wakeup with 20x 4.5 V/ 2 $\mu$s pulses, and low-voltage wakeup at 4.5V) after different numbers of 'set' pulses; \protect\subref{PWU_Gradual} gradual switching measurements for each pulse scheme, indicating the amount of switched polarization expected to be reached for the given pulse numbers
    \label{PWU_additional}
    }
\end{figure}

Figure \ref{FTJ_PWU} compares the FTJ device operation after several different pulsed wakeup schemes. In each case, the device operation pulses have the same parameters as those used for multi-pulse wakeup. The measurements were performed on a sample nominally identical to the previous results, although the overall current density was lower. Nonetheless, when considering the amount of domains contributing to switching after gradual switching experiments for each scheme, labelled in figure \ref{PWU_Gradual}, it is clear that the efficacy of a pulsed wakeup scheme can be estimated from gradual switching measurements employing the same pulse parameters. This provides a method for pre-selecting appropriate pulse parameters which can be used for both wakeup and device operation. 
\end{section}

\begin{section}{Additional comparison of pulsed wakeup and low-voltage wakeup}
\begin{figure}[h]
    \includegraphics[height=6cm]{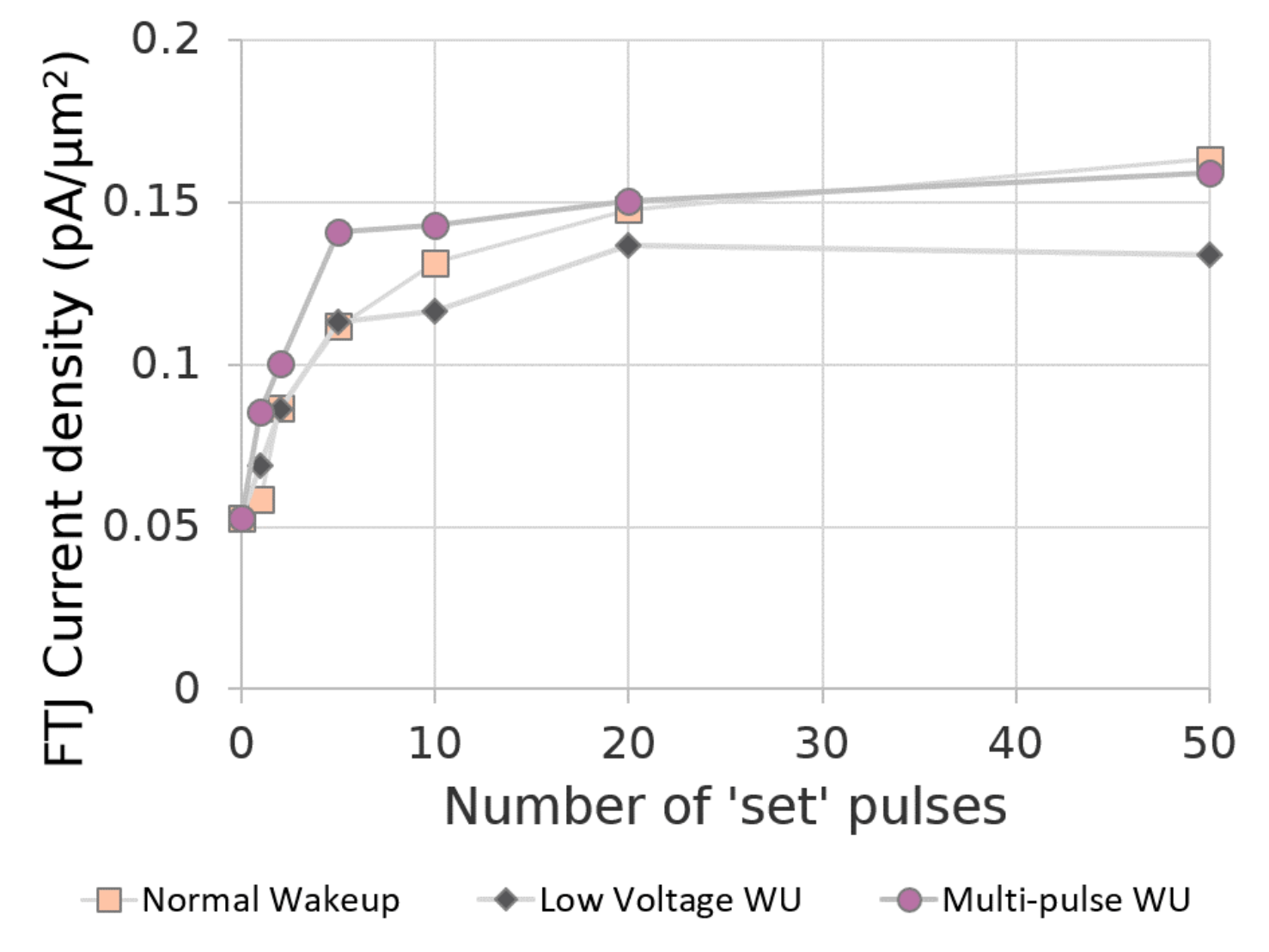}
    \caption{FTJ On-current measured after different wakeup schemes (normal wakeup, multi-pulse wakeup with 20x 4.5 V/ 2 $\mu$s pulses, and low-voltage wakeup at 4.5V) after different numbers of 'set' pulses
    \label{FTJ_4p5V}}
\end{figure}

A second parameter set was used to compare a multi-pulse wakeup (MPWU) scheme with a low-voltage wakeup (LVWU) scheme. In this case, 20x 4.5 V pulses of 2 $\mu$s were applied in each polarity. For the low-voltage wakeup, 40 $\mu$s, 4.5 V pulses were applied in each polarity. Again, the MPWU was more effective than LVWU, and the switching is faster for MPWU than for standard wakeup (6.5 V at 100 kHz). After LVWU, the device saturates at a lower on-current. 

\end{section}

\begin{section}{References}
[1] - H. Mulaosmanovic, T. Mikolajick, and S. Slesazeck, “Accumulative polarization reversal
in nanoscale ferroelectric transistors,” ACS applied materials \& interfaces 10, 23997–24002
(2018)
[2] - A. K. Saha, K. Ni, S. Dutta, S. Datta, and S. Gupta, “Phase field modeling of domain dynamics
and polarization accumulation in ferroelectric hzo,” Applied Physics Letters 114, 202903 (2019)
\end{section}